\documentclass[reprint,aps,showpacs,showkeys,prl,twocolumn]{revtex4-1}
\usepackage{graphicx,amssymb,amsmath,amsfonts,amsthm,hyperref}
\usepackage{algorithmic}
\newtheorem{theorem}{Theorem}
\newtheorem{algorithm}{Algorithm}
\newcommand{\ontop}[2]{\genfrac{}{}{0pt}{}{#1}{#2}}
\newcommand{\NN}{\mathbb{N}}
\newcommand{\ZZ}{\mathbb{Z}}

\newcommand{\PP}{\mathbb{P}}

\newcommand{\mix}{t_{\mathrm{mix}}} 
\newcommand{\TV}{\mathrm{TV}}
\newcommand{\C}{\mathcal{C}}
\newcommand{\sA}{\mathcal{A}}
\newcommand{\sE}{\mathcal{E}}
\newcommand{\sG}{\mathcal{G}}
\newcommand{\sL}{\mathcal{L}}
\newcommand{\sP}{\mathcal{P}}
\newcommand{\sY}{\mathcal{Y}}
\newcommand{\cycle}{\mathcal{C}_0}

\newcommand{\defect}{\mathcal{C}_2}
\newcommand{\doubledefect}{\mathcal{C}_4}
\newcommand{\worm}{\mathcal{W}}
\newcommand{\cp}{\sP}
\newcommand{\symdif}{\triangle}
\newcommand{\maxdeg}{\Delta}
\newcommand{\<}{\langle}
\renewcommand{\>}{\rangle}

\begin{document}

\title{The worm algorithm for the Ising model is rapidly mixing}
\date{\today}
\author{Andrea Collevecchio}
\affiliation{School of Mathematical Sciences, Monash University, Clayton, Victoria~3800, Australia}
\affiliation{University Ca'Foscari, San Giobbe, Cannaregio 873, 30121 Venezia Italy}
\author{Timothy M. Garoni}
\email{tim.garoni@monash.edu}
\affiliation{School of Mathematical Sciences, Monash University, Clayton, Victoria~3800, Australia}
\author{Timothy Hyndman}
\affiliation{School of Mathematical Sciences, Monash University, Clayton, Victoria~3800, Australia}
\author{Daniel Tokarev}
\affiliation{School of Mathematical Sciences, Monash University, Clayton, Victoria~3800, Australia}

\begin{abstract}
We prove rapid mixing of the Prokofiev-Svistunov (or \emph{worm}) algorithm for the zero-field ferromagnetic Ising model, on all finite
graphs and at all temperatures. As a corollary, we show how to rigorously construct simple and efficient approximation schemes for the Ising
susceptibility and two-point correlation function.
\end{abstract}
\maketitle

Markov-chain Monte Carlo (MCMC) simulation is one of the most versatile and widely-used tools applied in statistical physics. In order for
MCMC algorithms to be useful however, it is crucial that they converge rapidly to stationarity.

A major breakthrough in the development of efficient MCMC algorithms for statistical-mechanical spin models was the invention of the
Swendsen-Wang (SW) algorithm~\cite{SwendsenWang87}, which simulates the $q$-state Potts model~\cite{Potts52}. Careful numerical studies (see
e.g.~\cite{OssolaSokal04,DengGaroniMachtaOssolaPolinSokal07,GaroniOssolaPolinSokal11}) suggest that the SW algorithm can be considerably
more efficient than local single-spin flip algorithms. The SW algorithm utilizes a coupling~\cite{EdwardsSokal88} of the Potts and
Fortuin-Kasteleyn models~\cite{FortuinKasteleyn72,Grimmett06} to perform \emph{global} updates of the spins.

Recently however, it has been realized that some local algorithms have efficiencies comparable to, or even better than, the
SW algorithm. Indeed, recent numerical studies~\cite{DengGaroniSokal07_sweeny,ElciWeigel13} of the single-bond algorithm for the
Fortuin-Kasteleyn model, first studied by Sweeny~\cite{Sweeny83}, suggest that it is remarkably efficient, and exhibits the
surprising property of critical speeding-up~\cite{DengGaroniSokal07_sweeny}.

Another surprisingly efficient local algorithm is the \emph{worm} algorithm introduced by Prokofiev and
Svistunov~\cite{ProkofievSvistunov01}. Rather than simulating the original spin model, the Prokofiev-Svistunov (PS) algorithm simulates a
space of high-temperature graphs, using a clever choice of local moves. In~\cite{DengGaroniSokal07_worm}, a numerical study of the PS
algorithm concluded that it is the most efficient algorithm currently known for simulating the
susceptibility and correlation length of the three-dimensional Ising model. Numerical evidence presented in~\cite{Wolff09a} also suggests it
provides a very efficient method for studying the Ising two-point correlation function.

Despite the wealth of numerical evidence available for the SW, Sweeny and PS algorithms, relatively little is known rigorously
about the rate at which they converge to stationarity, or \emph{mix}. The SW algorithm is certainly the most well-studied of the
three: rapid mixing has been established at all non-critical temperatures on the square lattice~\cite{Ullrich12Thesis}, and the mixing of
the mean-field (complete graph) Ising case~\cite{LongNachmiasNingPeres14} has recently received a very careful treatment. Lower bounds on
the time required for mixing of the SW algorithm have also been established~\cite{LiSokal89,GoreJerrum99,BorgsChayesTetali12}. While no
rigorous results appear to have been established directly for the Sweeny algorithm, interesting comparison
results~\cite{Ullrich13,Ullrich14} have recently been proved which relate its mixing to that of the SW algorithm. To our knowledge,
no rigorous results have previously been reported for the PS algorithm.

In this Letter, we prove that the PS algorithm for the zero-field ferromagnetic Ising model is rapidly mixing, in a sense which we make
precise below. The result holds on all finite connected graphs, at all temperatures. In particular, it holds precisely at the critical point
on boxes in $\ZZ^d$. We are not aware of any other Markov chain for simulating the Ising model for which such a result is currently known.

As a corollary, we show how to rigorously construct simple and efficient approximation schemes for the Ising susceptibility and two-point
correlation function. Given the general nature of the methods used, we are optimistic that analogous arguments can be successfully applied
to PS algorithms for other models.

For an ergodic Markov chain with finite state space $\Omega$, transition matrix $P$, and stationary distribution $\pi$,
we define~\cite{LevinPeresWilmer09, Jerrum03} the \emph{mixing time} to be
\begin{equation} 
  \mix(\delta) := 
  \min\left\{ t \in \NN: \max_{s\in\Omega}\|P^t(s,\cdot)-\pi\|_{\TV} \le \delta\right\}
  \label{mixing time definition}
\end{equation}
where $\delta\in(0,1)$ and $\|\mu-\nu\|_{\TV}:=\max_{A\subseteq \Omega}|\mu(A)-\nu(A)|$ denotes the total variation distance between
measures $\mu$, $\nu$ on $\Omega$. The mixing time is therefore the first time the distribution of the chain comes within distance $\delta$
of stationarity, having started at a worst-possible initial state.

We say that a family of Markov chains, defined on state spaces of increasing size, is \emph{rapidly mixing} if $\mix(\delta)$ is bounded
above by a polynomial in $\log(|\Omega|)$. This implies that the chain need only visit a tiny fraction of the state space to ensure mixing,
so establishing rapid mixing is a very strong statement. For the Ising model, rapid mixing implies that although the number of
configurations is exponential in the number of sites, only a polynomial number of them need be visited to ensure mixing.

Consider now the ferromagnetic zero-field Ising model on finite connected graph $G=(V,E)$ at inverse temperature $\beta$. For any
$W\subseteq V$ and integer $1\le k \le |V|$ let
\begin{equation}
\C_W:=\{A\subseteq E:\partial A=W\}, \qquad \C_k := \bigcup_{\ontop{W\subseteq V}{|W|=k}}\C_W,
\label{configuration space}
\end{equation}
where $\partial A$ denotes the set of all odd vertices in the spanning subgraph $(V,A)$. Also consider
the unnormalized measure on $\{A\subseteq E\}$ defined by $\lambda(A) = x^{|A|}$ with $x=\tanh\beta$. The standard
high-temperature expansion of the Ising model~\cite{Thompson79} then gives the following graphical expression for the Ising correlation function
\begin{equation}
  \left\<\prod_{v\in W}\sigma_v\right\> = \frac{\lambda(\C_W)}{\lambda(\C_0)}.
  \label{high temperature expansion}
\end{equation}
The PS algorithm is defined on the configuration space $\worm=\cycle\cup\defect$, with stationary distribution
\begin{equation}
\pi(A) \propto x^{|A|} 
\begin{cases}
n, & A\in\C_0, \\
2, & A\in\C_2.
\end{cases}
\label{Ising to worm }
\end{equation}
The Ising susceptibility $\chi$ and two-point correlation function have natural expressions in terms of $\pi$
\begin{equation}
\chi = \frac{\beta}{\pi(\C_0)}, \qquad \<\sigma_u\sigma_v\> = \frac{n}{2}\frac{\pi(\C_{uv})}{\pi(\C_0)}.
\label{susceptibility and correlation wrt PS measure}
\end{equation}

A single step of the PS algorithm that we consider proceeds as in Alg.~\ref{PS algorithm}, with acceptance probabilities as given
in~\eqref{acceptance probabilities}.
\begin{algorithm}[PS algorithm]\hspace{\fill}\\
  \vspace{-5mm}
  \label{PS algorithm}
  \begin{algorithmic}
    \IF{$A\in\cycle$}
    \STATE Choose a uniformly random vertex $u\in V$
    \ELSIF{$A\in\defect$}
    \STATE Choose a uniformly random vertex $u\in\partial A$
    \ENDIF
    \STATE Choose a uniformly random neighbour $v$ of $u$
    \STATE With probability $a(A,A\symdif uv)$, let $A\to A\symdif uv$
    \STATE Otherwise $A\to A$
\end{algorithmic}
\end{algorithm}
\noindent Here $\symdif$ denotes symmetric difference. For technical reasons, we consider the \emph{lazy} version of the
algorithm, in which the acceptance probability is chosen to be one half of the standard Metropolis
prescription~\cite{Sokal97,LevinPeresWilmer09}
\begin{multline}
  \label{acceptance probabilities}
  a(A,A\symdif uv)
  = \\
  \frac{1}{2}
  \begin{cases}
    \displaystyle \min\left(1,\frac{d(u)}{d(v)}x^{\pm}\right), & A, A\symdif uv \in\defect, u\in\partial A\\
    \min(1,x^{\pm}), & \text{otherwise}.
  \end{cases}
\end{multline}
Here $x^{\pm}$ equals $x$ if the transition adds an edge, and $1/x$ if it removes an edge, and $d(u)$ denotes the degree of $u$ in $G$. If
$G$ is regular, then $a(A,A\symdif uv)$ is simply $\min(1,x^{\pm})/2$.

We now state our main result.
\begin{theorem}\label{mixing time bound}
  The mixing time of the PS algorithm on a finite connected graph $G=(V,E)$ with parameter ${x\in(0,1)}$ and $n=|V|\ge2$ satisfies
  $$
  \mix(\delta)\le\frac{1}{2x}\left(\log\left(\frac{8}{x}\right)-\frac{\log\delta}{m}\right)\!\left(3+\frac 1{mx}\right)\maxdeg(G)n^6m^2,
  $$
  where $m=|E|$ and $\maxdeg(G)$ is the maximum degree.
\end{theorem}
\noindent We note that general arguments imply that implicit in Theorem~\ref{mixing time bound} are bounds for other related properties of
the PS algorithm, including $O(\maxdeg(G)n^6m)$ bounds for the relaxation time (inverse spectral gap)~\cite{LevinPeresWilmer09}, exponential
autocorrelation time~\cite{Sokal97}, and all integrated autocorrelation times~\cite{Sokal97}. In the specific case of boxes in regular
lattices, each of these latter quantities are then $O(n^7)$, while $\mix(\delta)=O(n^8)$.

We now outline a proof of Theorem~\ref{mixing time bound}. A detailed proof will appear elsewhere~\cite{CollevecchioGaroniHyndmanTokarev14}.
Our argument uses the method of multicommodity flows~\cite{DiaconisStroock91,Sinclair92}. We therefore consider the \emph{transition graph}
$\sG$ of the PS algorithm, whose vertex set is $\worm$, and whose edge set $\sE$ consists of those pairs of
states $(A,A')\in\worm^2$ for which the one-step transition $A\to A'$ occurs with strictly positive probability. In its simplest form, the
method involves prescribing paths in $\sG$ between each pair of states $A,A'\in\worm$, and showing that for the given choice of paths there
are no edges in $\sG$ which become overly congested.

We now make these ideas precise. In the current setting, it is in fact convenient to define paths only from states in $\defect$ to states in
$\cycle$, rather than between all pairs in $\worm$. Therefore, for each pair $(I,F)\in \defect\times\cycle$, we fix a simple path
$\gamma_{I,F}$ in $\sG$ from $I$ to $F$, and we let $\Gamma=\{\gamma_{I,F}:(I,F)\in\defect\times\cycle\}$ denote the set of all such paths.
Adapting Lemma 4.4 from~\cite{JerrumSinclairVigoda04} to our setting implies that for any choice of $\Gamma$ we have
\begin{equation}
  \begin{split}
    \mix(\delta)
    &\le
    \log\left(\frac{1}{\pi_{\min}\delta}\right)
    \left[2+4\left(\frac{\pi(\defect)}{\pi(\cycle)}+\frac{\pi(\cycle)}{\pi(\defect)}\right)\right]\varphi(\Gamma)
    \\
    &\le
    \left(\log\left(\frac{8}{x}\right) - \frac{\log\delta}{m}\right)
    \left(6+\frac{2}{mx}\right) m\, n \, \varphi(\Gamma)
  \end{split}
  \label{JerrumSinclairVigodaLemma}
\end{equation}
where $\pi_{\min}=\min_{A\in\worm}\pi(A)$, and where the \emph{congestion} of $\Gamma$ is defined to be
$$
\varphi(\Gamma):=\sL(\Gamma) \max_{AA'\in\sE}\left\{\sum_{(I,F)\in\cp(AA')}\frac{\pi(I)\pi(F)}{\pi(A)P(A,A')}\right\}.
$$
Here $\cp(e)=\{(I,F)\in\defect\times\cycle:\ \gamma_{I,F}\ni e\}$ is the set of all pairs of states whose specified path uses the edge
$e\in\sE$, $\sL(\Gamma)=\max |\gamma_{I,F}|$ is the length of a longest path in $\Gamma$, and $P$ denotes the transition matrix of
the PS algorithm, as defined by Alg.~\ref{PS algorithm} and~\eqref{acceptance probabilities}. In obtaining the second inequality
in~\eqref{JerrumSinclairVigodaLemma} we have utilized the easily established bound~\cite{CollevecchioGaroniHyndmanTokarev14}
$$
\frac{2}{n}\frac{mx}{mx+1}\le\frac{\pi(\defect)}{\pi(\cycle)}\le n-1.
$$

The problem of bounding the mixing time has now been reduced to the problem of constructing an appropriate set of paths $\Gamma$ for which
tight bounds on the congestion $\varphi(\Gamma)$ can be obtained. We now exhibit such a set of paths. For concreteness, it is convenient to
fix some arbitrary vertex labeling of $G$, and to use this labeling to lexicographically induce an ordering on the set of all subgraphs of $G$.
For each cycle in $G$, we also use the vertex labeling to specify an arbitrary fixed orientation. 

\begin{figure*}[ht]
\begin{centering}
\includegraphics{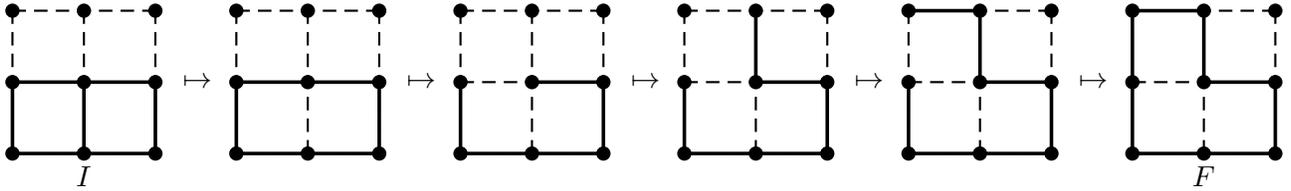}
\end{centering}
\caption{Example of a path $\gamma_{I,F}$. We order the vertices from left to right, and bottom to top. $I\symdif F=A_0\cup A_1$, where the path
  $A_0$ consists of the single edge $v_2v_5$, and the cycle $A_1$ is $v_4v_5v_8v_7v_4$.}
\label{paths figure}
\end{figure*}

We begin by noting that in order to transition from $I$ to $F$, it suffices to flip each edge in $I\symdif F$ precisely once. If
$(I,F)\in\C_2\times\C_0$, then $\partial(I\symdif F) = \partial I = \{u_I,v_I\}$ for some $u_I,v_I\in V$. By the handshaking lemma,
$u_I,v_I$ belong to the same component in $(V,I\symdif F)$. Let $A_0$ denote the shortest path between $u_I$ and $v_I$ in $(V,I\symdif F)$;
if multiple shortest paths exist, use the vertex labeling to ensure $A_0$ is uniquely defined. Now observe that $I\symdif
F\setminus A_0\in\C_0$. Since every element of the cycle space $\C_0$ can be decomposed~\cite{Diestel05} into an edge disjoint union of
cycles in $G$, we can again use the vertex labeling to obtain a unique decomposition $I\symdif F=\cup_{i=0}^k A_i$ for some $k$, where
$A_1,A_2,\ldots,A_k$ is an ordered list of disjoint cycles.

We can now define the path $\gamma_{I,F}$ as follows. Starting in state $I$, we first traverse the path $A_0$, starting from the lowest
labeled of the two odd vertices $\{u_I,v_I\}$, and inverting the occupation status of each edge as we proceed; add the edge if it was
absent, delete it if it was present. Having arrived at the intermediate state $I\symdif A_0$, we then process $A_1$, then $A_2$,\ldots. For
each cycle $A_i$, we begin at the lowest labeled vertex, and proceed according to the fixed orientation induced by the vertex labeling.
Once $A_k$ has been processed, we have arrived in state $I\symdif (\cup_{i=0}^kA_i)=F$. We emphasize that each step in the path
$\gamma_{I,F}$ corresponds to a valid step of the PS algorithm, which occurs with strictly positive probability. Let
$\Gamma=\{\gamma_{I,F}:(I,F)\in\defect\times\cycle\}$ denote the collection of all such paths.  Fig.~\ref{paths figure} illustrates a simple
example.

We now proceed to bound $\varphi(\Gamma)$ for this choice of $\Gamma$. Our argument is similar to that given in the discussion of
perfect and near-perfect matchings presented in~\cite{JerrumSinclairVigoda04}.  For each transition $e=AA'\in\sE$, we introduce a map
$\eta_e:\defect\times\cycle\to\worm\cup\doubledefect$ defined by $\eta_e(I,F):=I\symdif F\symdif A$. It is straightforward to
show~\cite{CollevecchioGaroniHyndmanTokarev14} that $\eta_e$ is injective. We also introduce the unnormalized measure $\Lambda$ on
$\worm\cup\doubledefect$ defined by
$$ 
\Lambda(A)=
x^{|A|}
\begin{cases}
  n, &\text{if $A\in \cycle$},\\
  2, &\text{if $A\in \defect$},\\ 
  1, &\text{if $A\in \doubledefect$}.
\end{cases}
$$
Note that for $A\in\worm$ we have $\pi(A) = \Lambda(A)/\Lambda(\worm)$.
It is again straightforward to show~\cite{CollevecchioGaroniHyndmanTokarev14} that
\begin{equation}
  \frac{\Lambda(I)\Lambda(F)}{\Lambda(A)} \le n\,\Lambda(\eta_e(I,F)).
  \label{Lambda bound}
\end{equation}
If $e=AA'$ is a maximally congested transition, then
\begin{align*}
  \varphi(\Gamma) 
  &\le \frac{m}{P(A,A')\Lambda(\worm)}n \sum_{(I,F)\in \cp(e)} \Lambda(\eta_e(I,F))\\ 
  &\le \frac{m n}{P(A,A')}\frac{\Lambda(\worm\cup\doubledefect)}{\Lambda(\worm)}\\ 
  &\le m n \frac{2n\maxdeg(G)}{x} \frac{n^3}{8} = \frac{1}{4x}\,\maxdeg(G)\,n^5\,m.
\end{align*}
The first inequality follows from~\eqref{Lambda bound} and the fact that $\sL(\Gamma)\le m$. The second follows because $\eta_e$ is an
injection. The third inequality then follows by noting that~\eqref{high temperature expansion} implies $\lambda(\C_W)\le\lambda(\C_0)$ for
any $W\subseteq V$, and also noting that~\eqref{acceptance probabilities} implies $P(A,A') \ge x/(2 n \maxdeg(G))$ for any $A\neq A'$ with
$P(A,A')>0$. This establishes Theorem \ref{mixing time bound}.

As immediate corollaries of Theorem~\ref{mixing time bound}, we can construct fully-polynomial randomised approximation schemes
(fpras)~\cite{KarpLuby89} for the Ising susceptibility and two-point correlation function. Both of these problems can be
shown~\cite{CollevecchioGaroniHyndmanTokarev14} to be \#P-hard, by reduction to the \#\textsc{Maxcut} problem, which is known to be
\#P-complete~\cite{JerrumSinclair93}. This strongly suggests that a general solution stronger than an fpras is unlikely to exist for these
problems.

Consider the susceptibility. An fpras for $\chi$ is a randomized algorithm such that for any $G$ and $\beta$, and any
$\epsilon,\eta\in(0,1/4)$, the algorithm runs in time bounded by a polynomial in $n,\epsilon^{-1},\eta^{-1}$, and the output $\sY$ satisfies
\begin{equation}
  \PP[(1-\epsilon) \chi \le \sY \le (1+\epsilon)\chi] \ge 1-\eta.
\end{equation}
From~\eqref{susceptibility and correlation wrt PS measure}, we see that in order to obtain an fpras for $\chi$, it suffices to
construct an fpras for $\pi(\cycle)$. 

Let $\sA\subseteq\worm$ be any event for which $\pi(\sA)\ge 1/S(n)$ with $S(n)$ a polynomial in $n$.  Let $R(G,\sA)$ denote the upper bound
for $\mix(\delta)$ given in Theorem~\ref{mixing time bound} with ${\delta=\epsilon/[16S(n)]}$.  A slight
refinement~\cite{CollevecchioGaroniHyndmanTokarev14} of Lemma 3 in~\cite{JerrumSinclair93} then implies that the following algorithm defines
an fpras for $\pi(\sA)$.
\begin{algorithm}[fpras]\hspace{\fill}\\
  \vspace{-5mm}
  \begin{algorithmic}
    \FOR{$1\le j \le 7 \lceil\log \eta^{-1}\rceil +1$}
    \FOR{$1\le i \le 20\,\lceil S(n)\epsilon^{-2}+1\rceil$}
    \STATE Run the PS algorithm for $R(G,\sA)$ steps
    \STATE Let $Y_{i,j}$ be 1 if the final state lies in $\sA$, and 0 otherwise
    \ENDFOR 
    \STATE Compute the sample mean $\overline{Y}_j$ of the $Y_{i,j}$
    \ENDFOR
    \STATE Output the median of $\{\overline{Y}_j\}$.
  \end{algorithmic}
  \label{fpras algorithm}
\end{algorithm}
\noindent Since it follows~\cite{CollevecchioGaroniHyndmanTokarev14} from~\eqref{high temperature expansion} that $\pi(\cycle)\ge 1/(2n+1)$, if in
Alg.~\ref{fpras algorithm} we let $\sA=\cycle$ and choose $S(n)=(2n+1)$, we obtain an fpras for $\pi(\cycle)$, and hence for $\chi$.

Similarly, fix $k\in\NN$, and consider the problem of computing the two-point correlation function between any pair of vertices $u,v$ of
graph distance ${d(u,v)\le k}$. Let $\sA=\C_{uv}$, let $S(n)=n(n+1)\,x^{-k}/2$, and note~\cite{CollevecchioGaroniHyndmanTokarev14} that
$\pi(\C_{uv})\ge 1/S(n)$. It then follows that Alg.~\ref{fpras algorithm} yields an fpras for $\pi(\C_{uv})$. Since we then have an fpras
for both $\pi(\C_{uv})$ and $\pi(\cycle)$, it follows from~\eqref{susceptibility and correlation wrt PS measure} that we have an fpras for
$\<\sigma_u\sigma_v\>$. To prove $\pi(\C_{uv})\ge1/S(n)$, note~\cite{CollevecchioGaroniHyndmanTokarev14} that for any fixed shortest path
$p_{uv}$ between $u$ and $v$, the map $\alpha:\C_{u,v}\to\cycle$ defined by $\alpha(A)=A\symdif p_{uv}$ is a bijection, which implies
$\lambda(\cycle)\le x^{-d(u,v)}\lambda(\C_{uv})$.

We conclude with some remarks. We note that Jerrum and Sinclair~\cite{JerrumSinclair93} also considered an MCMC algorithm on a space of
Ising high-temperature graphs, however their chain requires a strictly non-zero magnetic field. It can, nevertheless, be used to obtain an
fpras for the Ising partition function, even in zero field.

Finally, as noted in~\cite{DengGaroniSokal07_worm}, it is straightforward to establish a Li-Sokal type lower bound for the PS algorithm. In
particular, this implies that near criticality on $\ZZ_L^3$, the divergence of the relaxation time must be at least of order
$L^{d+\alpha/\nu}\approx n^{1.06}$, while Theorem~\ref{mixing time bound} implies it cannot be worse than $O(n^7)$. It would clearly be of
considerable interest to further sharpen these bounds in the specific setting of $\ZZ_L^3$, so as to determine the actual asymptotic
behaviour of the relaxation and mixing times in that case.

\begin{acknowledgments}
  The authors wish to gratefully acknowledge the contributions of Greg Markowsky to the early stages of this project.
  T.G. also gratefully acknowledges many fruitful discussions of the PS algorithm with Youjin Deng, Catherine Greenhill, Alan Sokal, Boris
  Svistunov and Ulli Wolff. This work was supported by the Australian Research Council (project numbers FT100100494, DP110101141 \&
  DP140100559). A.C. was also partially supported by STREP project MATHEMACS.
\end{acknowledgments}


%

\end{document}